\documentclass[aps,prl,twocolumn,showpacs]{revtex4}
\usepackage{amssymb}
\usepackage{graphicx}

\begin{document}

\title{Orbital Freezing in FeCr$_2$S$_4$ Studied by Dielectric Spectroscopy}

\author{R. Fichtl$^{1}$, P. Lunkenheimer$^{1}$, J. Hemberger$^{1}$, V. Tsurkan$^{1,2}$, and A. Loidl$^{1}$}

\address{$^{1}$Experimental Physics V, Center for Electronic Correlations and Magnetism,
University of Augsburg, 86135 Augsburg, Germany\\
$^{2}$Institute of Applied Physics, Academy of Sciences of
Moldova, Chisinau, R. Moldova}

\date{30.4.2004}

\begin{abstract}
Broadband dielectric spectroscopy has been performed on
single-crystalline FeCr$_2$S$_4$ revealing a transition into a
low-temperature orbital glass phase and on polycrystalline
FeCr$_2$S$_4$ where long-range orbital order is established via a
cooperative Jahn-Teller transition. The freezing of the orbital
moments is revealed by a clear relaxational behavior of the
dielectric permittivity, which allows a unique characterization of
the orbital glass transition. The orbital relaxation dynamics
continuously slows down over six decades in time, before at the
lowest temperatures the glass transition becomes suppressed by
quantum tunneling.
\end{abstract}


\maketitle

The cooperative and continuous freezing of translational and
orientational degrees of freedom into a highly degenerate glassy
state is one of the most fascinating phenomena in condensed matter
physics and far from being really understood \cite{Angell}.
Although canonical glass-forming materials and supercooled liquids
have been in the focus of scientific interest for about 200 years,
only in recent years the glassy dynamics has been investigated in
systems having perfect long-range translational order but being
disordered with respect to the orientational degrees of freedom.
Most prominent among those are spin glasses (SGs), featuring
diluted magnetic moments on a regular lattice. Disorder-derived
frustration suppresses any long-range magnetic order and the
moments cooperatively freeze into a glassy low-temperature state
\cite{Bind86}. Frustration in SGs results from the joint effect of
substitutional disorder and interactions that change sign as a
function of distance or bonding angle. Substitutional Mn ions in
Cu is a prototypical example. Glassy dynamics, can also be found
in a broad class of materials, including disordered
ferroelectrics, diluted molecular crystals, or ortho-para hydrogen
mixtures, which usually are termed orientational glasses (OGs)
\cite{Hoech90,Reger}. In OGs electric dipoles, elastic
quadrupoles, or even higher multipolar degrees of freedom
cooperatively freeze-in devoid of long-range orientational order.
Both, SGs and OGs are dominated by disorder-derived frustration
and exhibit a cooperative freezing of diluted moments on a regular
lattice.

Yet there exists another class of systems with frozen-in moments.
Namely geometrical frustration can establish complex ground states
with a large residual entropy, a problem that has been tackled
almost 50 years ago by Anderson \cite{Ande56} in treating the
magnetism of B-site spinel compounds. Geometrical frustration
arises if certain regular but specific arrays of interacting
moments can not satisfy all pair-wise interactions \cite{Ram01}.
Ising spins on a triangular lattice may serve as an illuminating
example. Spin ice, with a ground state derived from magnetic
moments located on a pyrochlore lattice, with ferromagnetic
interactions and a strong Ising-like anisotropy, is another
geometrically frustrated system, which recently gained
considerable interest \cite{RamNat}. Geometrical frustration has
been investigated in detail in the spin sector, resulting in most
cases in complex non-collinear spin structures or in spin-glass
behavior \cite{Ram01}. However, there exist recent experimental
reports on geometrical frustration also of the orbital moments
\cite{Frit04,Tsur04}. And while for FeSc$_2$S$_4$ the authors of
\cite{Frit04} propose a spin-orbital liquid, an orbital glass has
been detected in ferrimagnetic FeCr$_2$S$_4$ \cite{Tsur04}. The
possibility of orbital-glass (Jahn-Teller glass) formation has
been outlined long ago by Mehran and Stevens \cite{Mehr83}, and
the fact that orbital freezing can be detected by dielectric
spectroscopy has been demonstrated by Babinskii \textit{et al}.
\cite{Bab93}.

In canonical glass formers, dielectric spectroscopy has proven a
key technique for the investigation of glassy dynamics.
Especially, the exceptionally broad time/frequency window,
accessible with this method, makes it an ideal tool to follow the
many-decade change of molecular kinetics at the glass transition
\cite{Lunk00}. Here we show, that, as in FeCr$_2$S$_4$ the elastic
response of the ionic lattice is coupled to the orbital
reorientations via electron-phonon interaction, dielectric
spectroscopy also reveals valuable information on the orbital
freezing process. We provide a detailed dielectric investigation
of the orbital dynamics in single-crystalline FeCr$_2$S$_4$ and
compare it to results on polycrystalline material, which undergoes
a cooperative Jahn-Teller (JT) transition below 10 K. The slowing
down of the mean relaxation time, the distribution of relaxation
times, and the temperature dependence of the relaxation strength
are derived.

FeCr$_2$S$_4$ crystallizes in the normal cubic spinel structure
\cite{Frit04}. The Cr$^{3+}$ sublattice ($3d^{3}$, spin $S=3/2$)
is dominated by ferromagnetic exchange. The Fe$^{2+}$ ions
($3d^6$, $S=2$) are only weakly coupled within the Fe-sublattice,
but much stronger to the Cr ions. Hence FeCr$_2$S$_4$  reveals
ferrimagnetic spin order below $T_{c}=170$~K \cite{vStap82}. The
orbital moment of the octahedrally coordinated Cr$^{3+}$ ions is
quenched. Fe$^{2+}$ is tetrahedrally coordinated by the sulfur
ions; its lower e-doublet is occupied by three electrons and hence
is JT active. And indeed, in M{\"o}ssbauer experiments on
polycrystalline samples, an abrupt change of the electric-field
gradient below 10 K was attributed to a transition from a dynamic
to a static JT distortion, establishing an orbitally ordered
ground state \cite{Spen72}. In addition a $\lambda$-type anomaly
of the specific heat at 9.25 K was registered in Fe-deficient
samples \cite{Lotg72}. Recently it has been demonstrated that pure
single crystals reveal a transition into an orbital glass state,
while polycrystalline materials reveal long-range orbital order
\cite{Tsur04}.

Sample preparation and subsequent heat treatments of the poly-
(PC) and single-crystal (SC) samples are described in ref.
\cite{Tsur04}. The same SC and PC batches were investigated in the
present work. To characterize the samples, we measured the
magnetization in a commercial MPMS-5 SQUID magnetometer (Quantum
Design) and the thermal expansion in a home-built system utilizing
a $^4$He cryostat. For the dielectric measurements silver paint
contacts were applied to the plate-like samples forming a
parallel-plate capacitor. The conductivity and permittivity were
measured over a broad frequency range of 9 decades
($0.1~\mathrm{Hz} < \nu < 100~\mathrm{MHz}$) at temperatures down
to 1.4~K. A frequency response analyzer (Novocontrol
$\alpha$-analyzer) was used for frequencies $\nu<1$~MHz and a
reflectometric technique employing an impedance analyzer (Agilent
E 4291A) at $\nu>1$~MHz \cite{Schn01}.

\begin{figure} [b]
\includegraphics[width=6cm,clip]{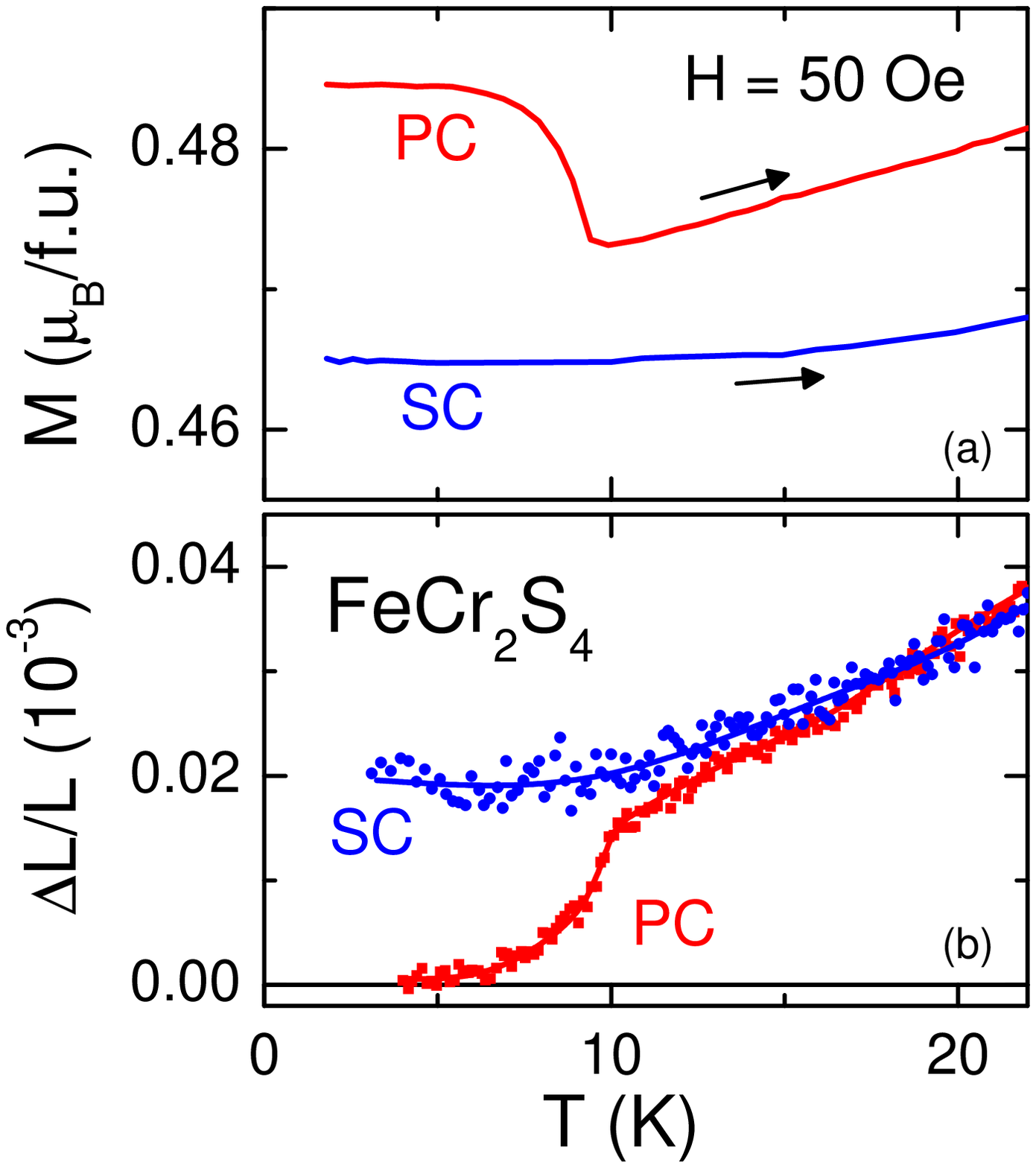}
\caption{(color online) Low-field (50~Oe) magnetization (a) and
thermal expansion (b) in ferrimagnetic FeCr$_2$S$_4$ SC and PC
samples.} \label{fig1}
\end{figure}

Figure \ref{fig1}(a) shows the field-cooled magnetization $M$ of
the PC and SC, measured in an external magnetic field of 50 Oe.
Only the PC reveals a upward jump of $M(T)$ just below 10 K,
indicative for a structural phase transition, while the SC shows
only a smooth temperature dependence. Also the thermal expansion
[Fig. \ref{fig1}(b)] provides evidence for a structural phase
transition only in the PC. These results are in accord with the
heat capacity results of Tsurkan \textit{et al}. \cite{Tsur04},
which reveal a Jahn-Teller transition at 9.2 K in the PCs, but a
clear indication of an orbital-glass transition in the SC. It is
thought that geometrical frustration dominates only in ideal
stoichiometric samples, but is partially broken by marginal
disorder, allowing for a cooperative phase transition restoring
orbital order. Such an order by disorder scenario has been treated
nearly 25 years ago by J. Villain \cite{Vill80}.

\begin{figure} [b]
\includegraphics[width=7.5cm,clip]{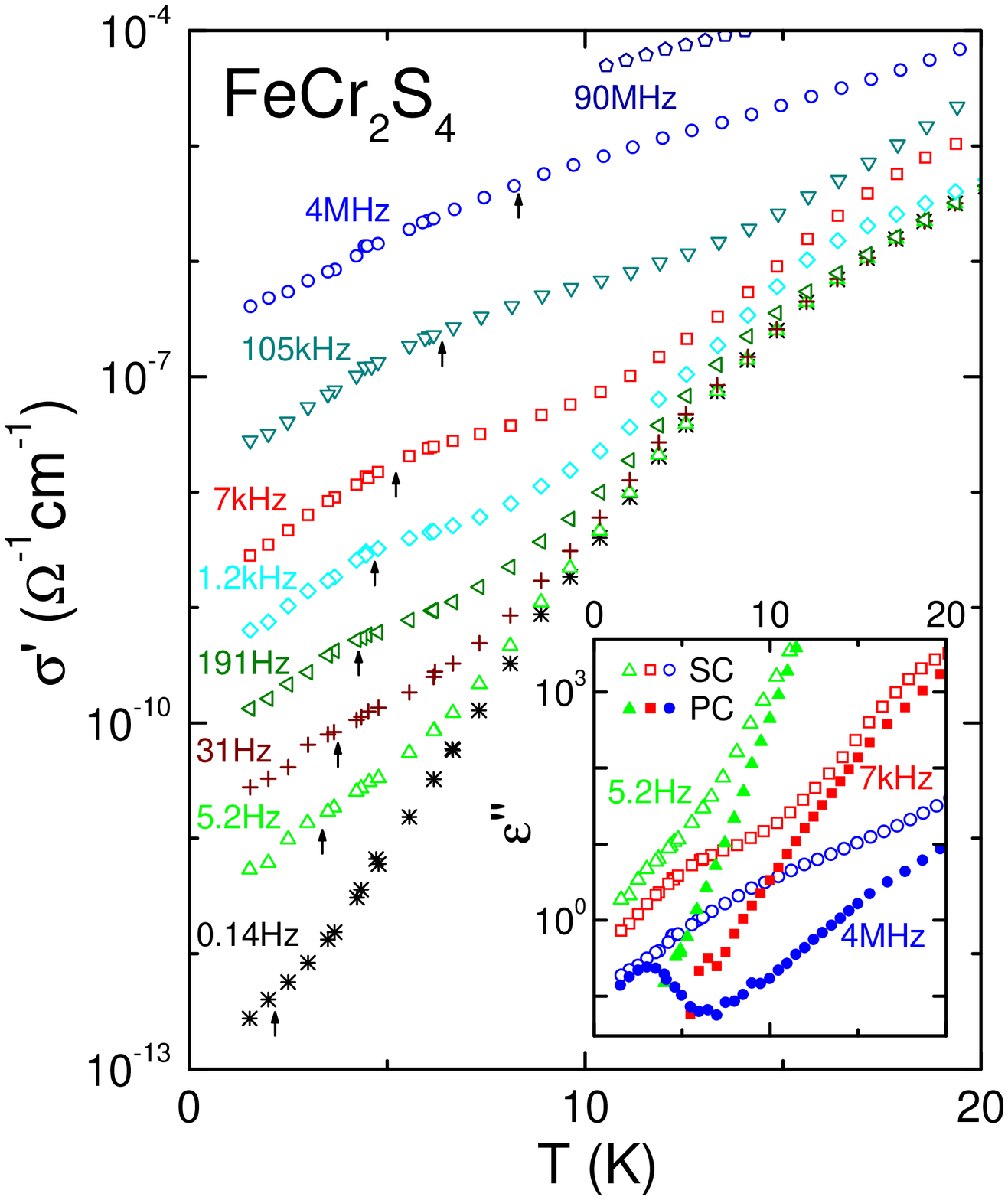}
\caption{(color online) Temperature dependence of the real part of
the conductivity in SC FeCr$_2$S$_4$ for various frequencies
between 0.14~Hz and 90~MHz. Inset: Comparison of the dielectric
loss in SC and PC material at selected frequencies.} \label{fig2}
\end{figure}

Figure \ref{fig2} shows the conductivity $\sigma^{\prime}$ of SC
FeCr$_2$S$_4$ for various frequencies at temperatures below 20~K.
The behavior is dominated by charge transport, namely a dc
contribution, approximately represented by the 0.14~Hz curve, and
an ac contribution increasing with frequency, which has a weaker
temperature dependence and can be ascribed to hopping transport of
localized charge carriers \cite{Elli87}. However, as indicated by
the arrows, superimposed to these contributions there is a
significant shoulder, indicating an underlying peak that shifts
towards lower temperatures with decreasing frequency. Having in
mind that
$\sigma^{\prime}\sim\varepsilon^{\prime\prime}\times\omega$, this
corresponds to a peak in the dielectric loss
$\varepsilon^{\prime\prime}(T)$, too. Thus, Fig. 1 reveals the
typical signature of relaxational behavior as commonly observed
e.g. for the glassy freezing of dipolar molecules
\cite{Hoech90,Lunk00}. In \cite{Tsur04} evidence for a glassy
freezing of the orbital dynamics was deduced from specific heat
measurements, displaying a cusp-like peak in $C_{p}/T$ vs. $T$ at
about 5~K. Thus it is reasonable that the relaxational feature in
$\sigma^{\prime}(T)$, occurring in just the same temperature
region, mirrors the glass-like slowing down of orbital dynamics.
Further evidence arises from a comparison to the results on the PC
sample (inset of Fig. \ref{fig2}), where the orbital degrees of
freedom are ordered at low temperatures \cite{Tsur04}. As
expected, the relaxation feature is absent in the PC,
$\varepsilon^{\prime\prime}$ being more than one decade smaller at
the peak temperature of the corresponding SC curve. In the PC, at
the highest frequency a small peak shows up at about 3~K, which,
however, is located at a significantly smaller temperature
compared with the shoulder observed at the same $\nu$ in the SC.
It also has a much smaller amplitude than the shoulder observed at
5.2~Hz in the SC, which is located at approximately the same
temperature, and thus its relaxational strength
$\Delta\varepsilon$ (corrected for the usual Curie-like behavior
$\Delta\varepsilon\sim1/T$) is nearly two decades smaller than for
the SC. We assume that defect centers in grain boundaries of the
Jahn-Teller ordered PC still undergo orbital reorientations. From
the relaxation strength we estimate this fraction of non-ordered
arrays to be of the order of 1\%.

\begin{figure} [b]
\includegraphics[angle=270,width=8cm,clip]{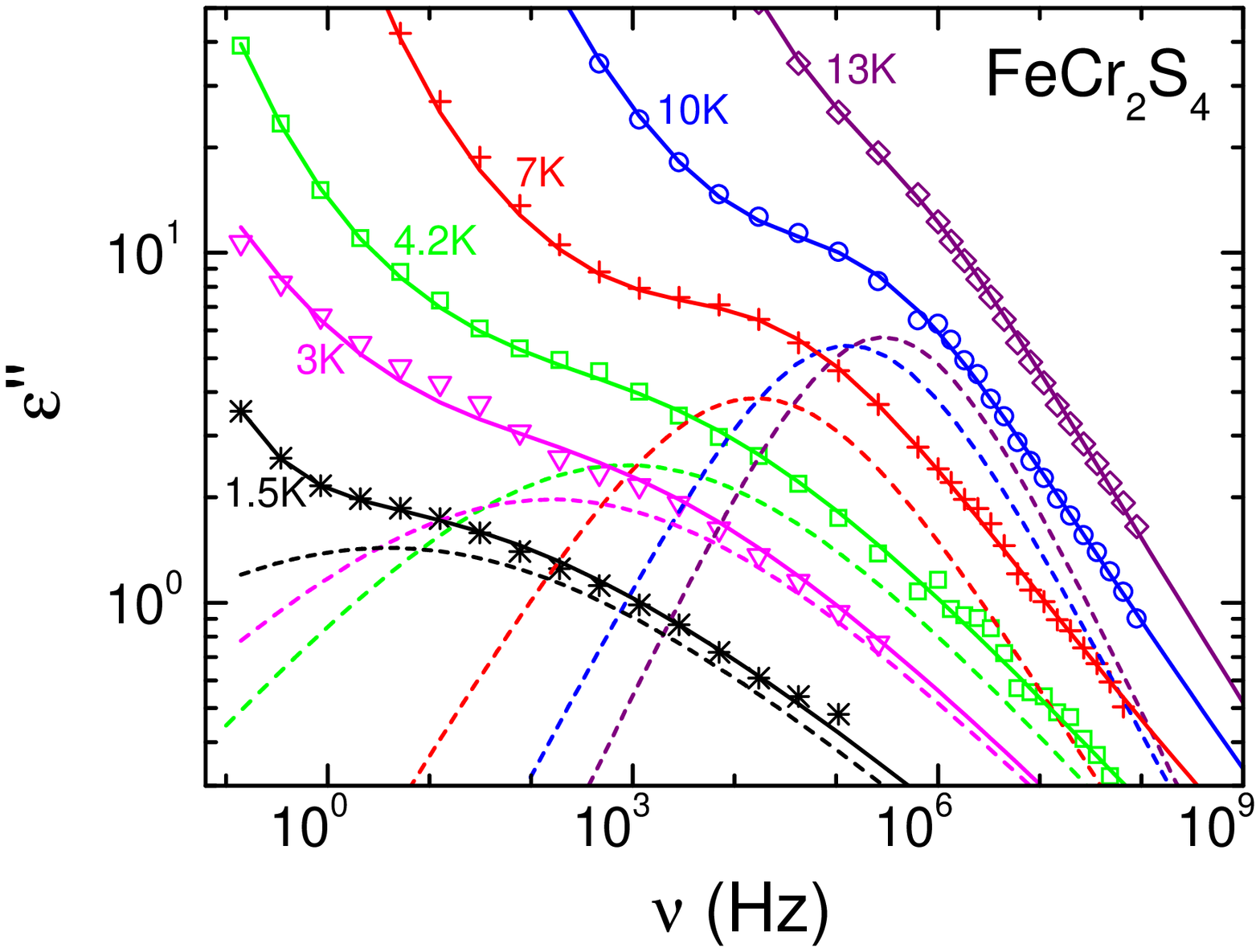}
\caption{(color online) Dielectric loss \textit{vs.} frequency at
temperatures $T\leq 13$~K. The solid lines represent the results
of fits as described in the text. The dashed lines characterize
the contributions due to the orbital relaxations.} \label{fig3}
\end{figure}

In the frequency-dependent plot of Fig. \ref{fig3}, the loss peaks
characterizing the orbital relaxation are analyzed in more detail.
The solid lines represent fits with the sum of a Cole-Cole (CC)
function \cite{Cole41}, often employed to describe loss peaks in
canonical and OGs, and a conductivity contribution,
$\sigma^{\prime}=\sigma_{dc}+\sigma_{0}\omega^{s}$, leading to a
divergence of $\varepsilon^{\prime\prime}\sim\sigma^{\prime}/\nu$
towards low $\nu$. The latter is composed of a dc component and an
ac power-law contribution with exponent $s<1$, representing the
so-called "universal dielectric response" (UDR) \cite{Jons83}. For
semiconducting systems, the UDR is the signature of hopping
conduction of Anderson-localized charge carriers and explained
within various theoretical approaches \cite{Elli87}. Good
agreement of fits and experimental spectra could be achieved in
this way, the dashed lines showing the relaxational part of the
fits. Judging these results, one should bear in mind their rather
high uncertainty, due to a partial correlation of the fit
parameters. Whatsoever, as can be deduced even from the raw data,
the loss peaks must be broader than for the Debye case,
corresponding to a single relaxation time. Thus the relaxation in
FeCr$_2$S$_4$ shows the typical broadening of glassy systems,
commonly ascribed to a heterogeneous distribution of relaxation
times. With decreasing temperature the loss peak broadens
significantly, while its amplitude decreases. Such a narrowing of
a CC-like loss peak is commonly observed in OGs. It can be
explained assuming a temperature-independent Gaussian distribution
of energy barriers, leading to loss peaks whose widths increase
like 1/T \cite{gauss}.

\begin{figure} [b]
\includegraphics[angle=270,width=8cm,clip]{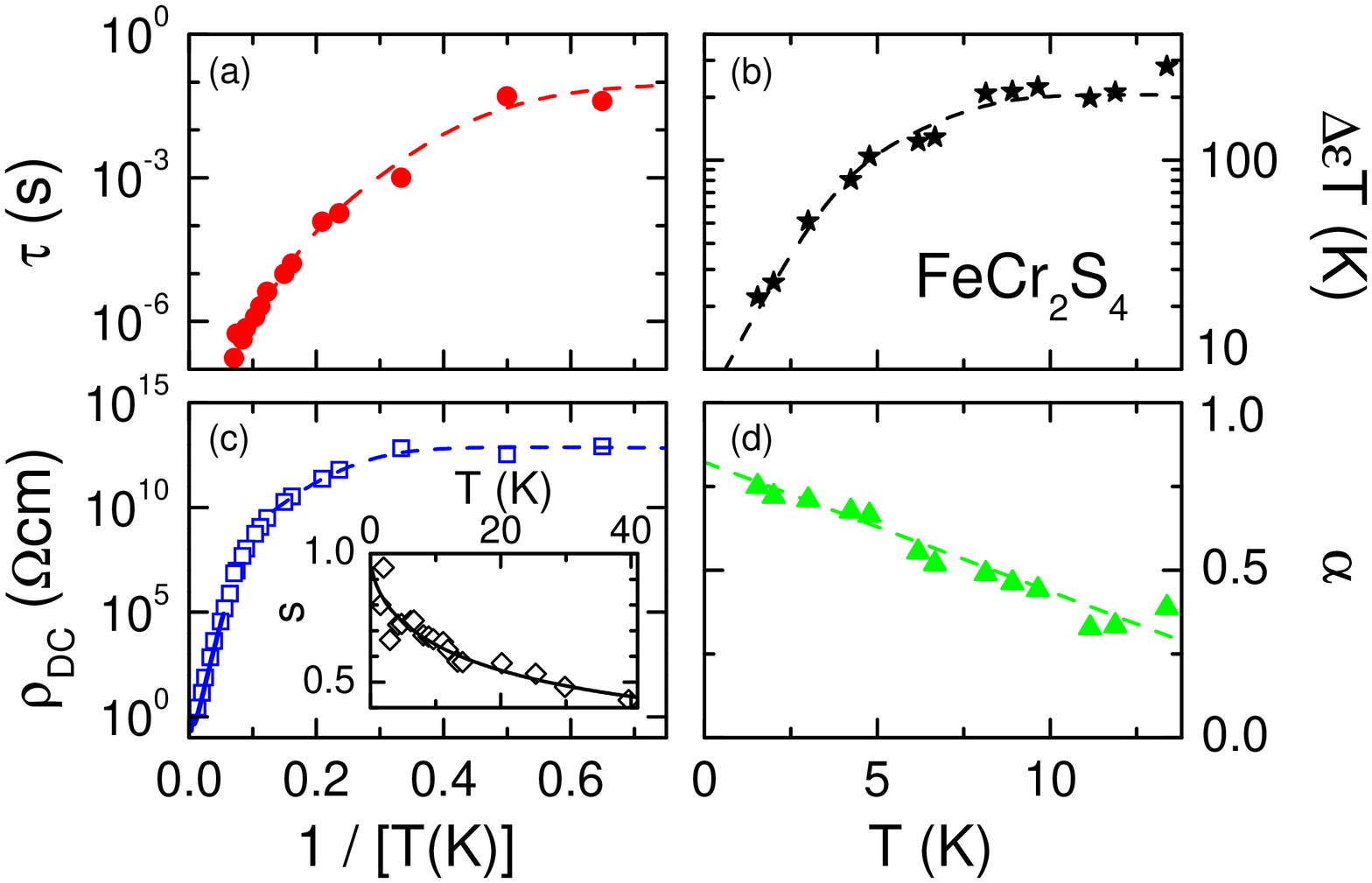}
\caption{(color online) Temperature dependence of relaxation time
(a), relaxation strength times T (b), dc resistivity (c), and
width parameter of the CC distribution (d) as resulting from the
fits shown in Fig. \ref{fig3}. The dashed lines are guides to the
eyes. The solid line in (c) shows the result of a four-point
dc-measurement. In the inset of (c) the power-law exponent $s$ is
given; the line is a fit with a polaron tunneling model
\cite{Elli87}.} \label{fig4}
\end{figure}

In Fig. \ref{fig4} the temperature dependences of the main fit
parameters are given. The relaxation time $\tau$ characterizing
the reorientational dynamics of the orbitals [Fig. \ref{fig4}(a)],
shows a smooth variation over many decades, which is typical for
glassy freezing. However, in contrast to most other glassy
systems, its temperature dependence becomes weaker for low
temperatures. In the Arrhenius-representation, $\log(\tau)$
\textit{vs}. $1/T$, a purely thermally activated process
corresponds to a straight line, while tunneling would yield a
nearly constant $\tau(T)$. Obviously Fig. \ref{fig4}(a) represents
a smooth transition from a thermally activated ($T>20$~K) to a
purely tunneling type ($T<2$~K) of reorientation. That indeed
quantum fluctuations can suppress the glass transition has been
demonstrated theoretically for a proton glass \cite{Dobr87}. In
Fig. \ref{fig4}(b), the relaxation strength $\Delta\varepsilon$
multiplied by temperature is plotted. For $T>5$~K,
$\Delta\varepsilon\times T$ is constant, which indicates that
$\Delta\varepsilon$ approximately follows a Curie-type behavior,
characteristic for non-interacting orbitals. Below about 5~K
$\Delta\varepsilon\times T$ starts to decrease on decreasing
temperature signaling a static freezing of a significant fraction
of reorienting orbitals, which are then lost for the relaxation
process. It indicates a type of smeared-out phase transition with
a certain number of orbitals fixed in random orientations while
the remaining ones undergo a reorientational motion via tunneling
processes. From Fig. \ref{fig4} important differences of the
orbital relaxation dynamics can be derived when compared to
canonical glass formers. In supercooled liquids, the slowing down
of the reorientational motion is stronger than an Arrhenius law
and the relaxation peaks are significantly asymmetric. However, in
FeCr$_2$S$_4$ the orbital motion slows down even weaker than
Arrhenius because quantum tunneling suppresses conventional
freezing. Fig. \ref{fig4}(d) shows the width parameter $\alpha$ of
the CC function ($\alpha>0$ corresponds to symmetrically broadened
peaks, compared to Debye behavior with $\alpha=0$). $\alpha$
increases linearly with decreasing temperature, which simply
reflects the fact that the relaxational motion is dominated by a
Gaussian distribution of energy barriers \cite{gauss}. Based on a
linear extrapolation of $\alpha(T)$ in Fig. \ref{fig4}(d), the
relaxation is monodispersive and Debye-like for temperatures
$T>20$~K and the distribution of relaxation times is almost
infinitely broad towards $T=0$~K. Based on Fig. \ref{fig4}(b) one
has to bear in mind that a significant fraction of orbitals no
longer takes part in the relaxational process due to cooperative
static freezing. This quenched disorder, resulting from frozen-in
orbitals and local strain-fields, yields locally varying energy
barriers and a symmetric and broad distribution of relaxation
times.

In Fig. \ref{fig4}(c) the dc resistivity $\rho_{DC}$ as deduced
from the fits is given. It agrees perfectly with the results of a
four-point dc-measurement (solid line), performed at $T>19$~K. At
$T\gtrsim14$~K), $\rho_{DC}(T)=1/\sigma_{dc}(T)$ behaves thermally
activated with an energy barrier of about 23~meV. At lower
temperatures, the temperature dependence of $\rho_{DC}$ becomes
successively weaker, being nearly constant for $T<3$~K, indicating
charge transport via tunneling. The inset of Fig. \ref{fig4}(c)
shows the power law exponent $s(T)$ of the hopping-dominated ac
conductivity. It increases approaching unity for low temperatures
and can be described within a model for hopping transport via
polaron tunneling (solid line) \cite{Elli87}. Thus there is
evidence for tunneling-dominated dynamics for both, the
reorientational motion of the orbitals and the charge transfer
from site to site. Interestingly, the tunneling regime seems to be
identical for $\tau$ and $\rho_{DC}$ and the curves of Figs.
\ref{fig4}(a) and (c) can be scaled to nearly perfectly match each
other. We interpret this finding assuming that both, the
reorientation of the electronic charge distribution and the
hopping of electrons from site to site, via strong electron-phonon
coupling are accompanied by the same kind of structural
rearrangement, which determines the temperature evolution of the
corresponding quantities. For the charge transport this is
corroborated by the indication of polaronic charge carriers
deduced from $s(T)$.

In conclusion, by dielectric spectroscopy at low temperatures, we
achieved a thorough characterization of the glassy freezing of the
orbital degrees of freedom in FeCr$_2$S$_4$. In contrast to most
glassy materials, this low-temperature state is not established
via randomness, but most probably is governed by geometrical
frustration. We find typical glassy behavior, in particular a
continuous slowing down of the orbital dynamics and a distribution
of relaxation times, which almost diverges for $T\rightarrow 0$~K.
Based on the temperature dependence of $\Delta\varepsilon$, a
large fraction of orbitals undergoes a static freezing, while for
the remaining entities a conventional freezing-in is suppressed by
quantum-mechanical tunneling, limiting the low-temperature
relaxation time to $\tau\simeq 10^{-1}$~s. The low-temperature
charge transport in FeCr$_2$S$_4$ is dominated by tunneling of
polaronic charge carriers.

This work was supported by the Deutsche Forschungsgemeinschaft via
the Sonderforschungsbereich 484 and partly by the BMBF via
VDI/EKM, FKZ 13N6917.

\end{document}